\documentclass[journal]{IEEEtran}

\IEEEoverridecommandlockouts    

\usepackage{enumitem}
\usepackage{graphicx}
\usepackage{algorithmic}
\usepackage{epsfig}
\usepackage{wrapfig}
\usepackage{amsmath,amsthm, amssymb}
\usepackage[]{algorithm2e}
\usepackage{multirow}
\usepackage{multicol}
\usepackage{tabularx}
\usepackage[dvipsnames]{xcolor}
\usepackage{comment}
\usepackage{subcaption}

\usepackage{array}
\usepackage{color}
\usepackage{breqn}
\usepackage{cite}
\usepackage{booktabs}

\newcommand{\N}{\mathcal{N}}
\newcommand{\E}{\mathcal{E}}
\newcommand{\G}{\mathcal{G}}

\newcommand{\A}{\mathcal{A}}

\begin{document}

\title{Simulation-Integrated Distributed Optimization for Unbalanced Power Distribution Systems}

\author{Rabayet~Sadnan~\IEEEmembership{Student Member,~IEEE}, Nathan~Gray~\IEEEmembership{Student Member,~IEEE}, ~Anjan~Bose,~\IEEEmembership{Senior Member,~IEEE}   and~Anamika~Dubey,~\IEEEmembership{Senior Member,~IEEE}
\vspace{-20pt}
\thanks{Rabayet Sadnan, Nathan Gray, Anjan Bose and Anamika Dubey are with the Department of EECS, WSU, Pullman, WA-99164, USA; e-mail: rabayet.sadnan@wsu.edu}%
}




\maketitle

\begin{abstract}
Distributed optimization methods have been extensively applied for the optimization of electric power distribution systems, especially for grid-edge coordination. Existing distributed optimization algorithms applied to power distribution systems require many communication rounds among the distributed agents and may pose convergence challenges in difficult nonlinear settings. The communication network parameters also significantly impact the algorithm's performance. In this paper, we propose a scalable, equivalent network approximation-based, distributed optimization algorithm that employs \textit{simulation within optimization} using the system’s digital twin (DT) to solve the optimal power problems (OPF) for a three-phase unbalanced distribution system. The proposed approach is implemented using a cyber-physical co-simulation platform to validate the robustness of the proposed distributed algorithm under stressed communication. The proposed approach is thoroughly validated using the IEEE 123-bus test system.

\end{abstract}

\IEEEpeerreviewmaketitle

\vspace{-0.3cm}
\vspace{-0.4cm}
\section{Introduction}
Optimization of power distribution systems has gained significant attention with growing penetrations of controllable distributed energy resources (DERs). Traditionally, the distribution system is centrally managed using an advanced distribution management system (ADMS) that collects the measurement data and use it to coordinate DERs and other grid-edge technologies. With the increasing number of such controllable assets and growing system complexity, a centrally managed system suffers from several challenges including poor scalability, and susceptibility to a single-point failures \cite{momoh1999review1,castillo2013survey,molzahn2017survey}. The additional nonlinearity introduced by the three-phase and unbalanced power distribution systems may pose convergence issues for the centralized optimization problems and may result in failure to attain the optimal solution within a reasonable time \cite{molzahn2017survey}. Besides, a centrally managed system is impractical for emerging distribution system typically composed of several geographically separated areas controlled by different entities that may not want to share model and/or data. To address these challenges, distributed coordination and optimization paradigm has gained significant attention for grid-edge coordination. 

The existing literature has shown the applicability of the distributed optimization paradigm in reducing the computational requirements for optimal power flow (OPF) problems and in mitigating other challenges including ownership boundaries and avoiding single-point failures \cite{molzahn2017survey}. To this end, general distributed optimization techniques, such as, Alternating Direction Method of Multipliers (ADMM), Auxiliary Problem Principle (APP), etc. have been adopted to solve distributed OPF problems for power distribution systems \cite{ kim2000comparison,zheng2015fully,boyd2011distributed}. Distributed OPF algorithms involve two iterative processes: (1) sub-problem optimization involving micro-iterations, where distributed agents individually solve optimization problems specific to their areas, (2) distributed coordination requiring macro-iterations, where distributed agents exchange the boundary variables typically with their immediate neighbors. The overall compute complexity of the algorithm depends upon the total number micro- and macro- iterations taken to achieve network-level convergence. The nature of the optimization problem dictates micro-iterations needed to solve each sub-problem; the added nonlinearity with three-phase OPF may pose convergence and compute challenges for this step. The distributed coordination approach dictates the number of macro-iterations needed for network-wide convergence; traditional algorithms take 100s of rounds of communication exchange or macro-iterations resulting in a slow converge. This calls for algorithmic advances in distributed OPF algorithms to realistically apply them to a real-world distribution system. 

Although several recent methods attempted to reduce the computational requirements for local subproblems \cite{qu2017harnessing,mhanna2018adaptive,peng2016distributed}, challenges with convergence and speed persist due to added nonlinearities in three-phase unbalanced OPF models for power distribution system. Earlier {\em optimization with simulation} has been introduced for general complex optimization problems that can leverage the already developed complex models of the system, and avoid including the detailed system model in the optimization problem formulation \cite{april2004new, nakayama2002simulation,nguyen2014review}. Although this approach has not been directly applied to solve OPFs for power systems, the use of high-fidelity simulation models for design optimization has been reported \cite{zhou2019digital,pan2020digital}. This high-fidelity simulation model is referred as a digital twin (DT) of the physical power system and has been employed to aid with situational awareness and decision-making for power grid operations  \cite{brosinsky2018recent, he2019preliminary, darbali2021state}. The DT models provide an ability to perform high-fidelity simulations of the grid that can potentially help reduce the complexity of difficult OPF problems. However, the use of DT models has been limited to evaluating the physical system in advance for potential actions; their utility in directly assisting with difficult power system optimization problems has not been investigated.

Another challenge with traditional algorithms to solve distributed OPF algorithms emerge from a large number of macro-iterations ($\ge 100$ iterations) needed to converge for relatively small and/or balanced systems \cite{erseghe2014distributed, dall2013distributed,molzahn2017survey}.  Thus, these algorithms require an expensive communication infrastructure with high bandwidth and low communication delays. Although real-time feedback-based online distributed algorithms address some of the challenges \cite{bernstein2019real,magnusson2020distributed}, they still require thousands of time-steps to track the optimal solution, resulting in a sub-optimal system operation and constraint violations. Recently, authors proposed an algorithm based on the structure of the power flow problem for a radial distribution system that drastically reduces the number of macro-iterations  \cite{ sadnan2021distributed }. However, the sub-problems are modeled as nonlinear programming (NLP) problems that may pose convergence and compute challenges. Moreover, the performance of distributed OPF algorithms has not been evaluated in a {\em realistic cyber-physical environment} with appropriate models for  communication systems ~\cite {sadnan2021distributed, erseghe2014distributed,magnusson2020distributed}.

The objective of this paper is to develop a fast, computationally tractable, and communication-efficient distributed OPF (D-OPF) algorithm for {\em{unbalanced}} power distribution system. Our approach employs {\em optimization with simulation} concepts using DTs (at sub-problem level) and a distributed coordination algorithm to obtain a computationally efficient and near-optimal D-OPF approach. Specifically, we optimize a low-fidelity linear model of the sub-system and project the obtained solution to the high-fidelity simulation model using DTs resulting in a near-optimal but feasible power flow solution. Primarily, the use of DTs (at sub-problem level) excludes the necessity of solving complex nonlinear OPFs and reduces the overall solution time for the local subproblems. Then the distributed coordination algorithm is employed to exchange the projected power flow variables with the neighboring distributed areas. We carefully select shared variables based on the power flow in radial power distribution systems that help us reduce the required number of macro-iterations by an order of magnitude. Together these innovations result in highly scalable and near-optimal algorithms for solving difficult D-OPF problems.  The proposed approach is also extensively evaluated using a comprehensive cyber-power co-simulation environment.

\vspace{-0.3cm}
\section{Network Modeling \& Problem Formulation}
In this paper, $(\cdot)^T$ represents matrix transpose; $(\cdot)^*$ represents the complex-conjugate; $(\cdot)^{(n)}$ represents the $n^{th}$ iteration;  $\mathbb{Re}$ denotes the real part of the complex number; the superscript $p$ (without parenthesis) denotes the three-phases, i.e., $\{a,b,c\}$ of the system. 
\vspace{-0.3cm}
\subsection{Power Flow Model}
\vspace{-0.1cm}
This section details the power flow models to formulate the OPF problems for an unbalanced power distribution system. The unbalanced power flow equations are based on our prior work \cite{jha2019bi}. Let us consider an unbalanced radial power distribution network of $n$ buses where, $\N$ denotes the set of buses in that system and $\E$ denotes the set of edges identifying distribution lines that connect the ordered pair of buses $\{ij\}$, $\forall$ $i,j \in \N$. Here, $\phi_j$ denotes the set of phases in the bus $j$. Let $v_j^p =|V_j^p|^2$ be the squared magnitude of voltage at bus $j\in \N$ for phase $p\in \phi_j$. Define $\phi_{ij} = \{pq : p\in \phi_i$ and $q\in \phi_j$  $\forall \{ij\}\in \E\} $. Let $l_{ij}^{pq} =(|I_{ij}^p||I_{ij}^q|)$ be the squared magnitude of the line current flowing in the phase ${pq}\in \phi_{ij}$ of line $(i,j)$. Also, $S^{pq}_{ij}=P^{pq}_{ij}+ Q^{pq}_{ij}$ and $z^{pq}_{ij}=r^{pq}_{ij}+j x^{pq}_{ij}$, where $pq \in \phi_{ij}$. Variable $p_{L,j}^p$ and $q_{L,j}^p$ denote the active and reactive load (respectively) connected at node $j$ of phase $p\in \phi_j$. Similarly, subscript $D,j$ denotes the DER generation at node $j$; e.g., $p^p_{D,j}$ denotes the active power generation at node $j$ for phase $p$. $\delta^{pq}_{ij}$ is the angle difference between the phase currents. First, we define the nonlinear power flow model and then the linearized model is detailed. For more details, please refer to \cite{jha2019bi}.

\subsubsection{Nonlinear Model}
With the approximated phase voltage and branch current angles, in \cite{jha2019bi} developed the nonlinear power flow model for an unbalanced radial power distribution system. The model is defined below in \eqref{nl_pf}. We model the loads in the systems as constant power loads.

\vspace{-0.4cm}
\begin{small}
\begin{IEEEeqnarray}{C C}
\IEEEyesnumber\label{nl_pf} \IEEEyessubnumber*
\nonumber  P_{ij}^{pp} - \sum_{q \in \phi_j}{l_{ij}^{pq} \left(r_{ij}^{pq} \cos(\delta_{ij}^{pq})- x_{ij}^{pq} \sin(\delta_{ij}^{pq})\right)}\\ 
   =\sum_{k:j \rightarrow k}P_{jk}^{pp} + p_{L,j}^p- p_{D,j}^p\\
    \nonumber  Q_{ij}^{pp} - \sum_{q \in \phi_j}{l_{ij}^{pq}\left(x_{ij}^{pq} \cos(\delta_{ij}^{pq})+ r_{ij}^{pq} \sin(\delta_{ij}^{pq})\right)}\\
 = \sum_{k:j \rightarrow k}Q_{jk}^{pp}+q_{L,j}^p  - q_{D,j}^p - q_{C,j}^p \\
   \nonumber v_j^p = v_i^p - \sum_{q \in \phi_j}{2 \mathbb{Re}\left[S_{ij}^{pq} (z_{ij}^{pq})^*\right]} + \sum_{q \in \phi_j}{z_{ij}^{pq} l_{ij}^{qq}}\\
   +\sum_{q1,q2 \in \phi_j, q1 \neq q2}{2\mathbb{Re}\left[ z_{ij}^{pq1} l_{ij}^{q1q2}\left(\angle(\delta_{ij}^{q1q2})\right)(z_{ij}^{pq2})^*\right]}\\
   (P_{ij}^{pp})^2 + (Q_{ij}^{pp})^2 = v_i^p  l_{ij}^{pp}\\
   (l_{ij}^{pq})^2 = l_{ij}^{pp}  l_{ij}^{qq}
\end{IEEEeqnarray}
\end{small}
\vspace{-0.45cm}

\subsubsection{Linear-approximated Model}
The computational complexity augmented by the nonlinear power flow models to the existing scalability issues associated with large-scale OPF problems can be reduced using the approximated linearized power flow models. The linearized power flow models can solve the power flow variables, such as nodal voltages, within reasonable accuracy and reduce the computation time by the order of magnitudes making the linear models scalable for large-scale power distribution systems. A linear-approximated power flow model, three-phase LinDistFlow model, is defined in \eqref{lin_pf}. The primary assumption is that the line loses are negligible compared to the power flow in the system. 

\vspace{-0.4cm}
\begin{small}
\begin{IEEEeqnarray}{C C}
\IEEEyesnumber\label{lin_pf} \IEEEyessubnumber*
P_{ij}^{pp} =\sum_{k:j \rightarrow k}P_{jk}^{pp} + p_{L, j}^p- p_{D, j}^p\\
Q_{ij}^{pp} = \sum_{k:j \rightarrow k}Q_{jk}^{pp}+q_{L,j}^p  - q_{D,j}^p - q_{C,j}^p \\
v_j^p = v_i^p - \sum_{q \in \phi_j}{2 \mathbb{Re}\left[S_{ij}^{pq} (z_{ij}^{pq})^*\right]}
\end{IEEEeqnarray}
\end{small}
\vspace{-0.85cm}

\subsection{DER Modeling}
We define the DER models for different network objectives. In this paper, to achieve minimal cost, either reactive or active power generation of the DERs is set as the decision variables of the OPF problems. The active power generation, $p^p_{D,j}$ is assumed to measured and known, when the reactive power generation, $q^p_{D,j}$ is the decision variable. If $S^p_{DR,j}$ is defined as the nominal rating of the DER at node $j$ for phase $p\in \phi_j$, then the DER is modeled as \eqref{DER_Q}.

\vspace{-0.3cm}
\begin{small}
\begin{equation}\label{DER_Q}
-\sqrt{(S_{DR,j}^{p})^2 - (p_{D,j}^p)^2} \leq q_{D,j}^p \leq \sqrt{(S_{DR,j}^{p})^2 - (p_{D,j}^p)^2} 
\end{equation}
\end{small}
\vspace{-0.4cm}

For network objectives, such as DER maximization, the active power generation, $p^p_{D,j}$ is set as decision variables, and we assume $q^p_{D,j} = 0$. The DER model is defined in \eqref{DER_P}. However, if both active and reactive power generation is considered as decision variables, then the DER is modeled as a second-order conic equation as defined in \eqref{DER_PQ}.

\vspace{-0.2cm}
\begin{small}
\begin{equation}\label{DER_P}
\small
0 \leq p_{D,j}^p \leq S_{DR,j}^{p}
\end{equation}
\vspace{-0.4cm}
\begin{equation}\label{DER_PQ}
\small
\left(p_{D,j}^p\right)^2+ \left(q_{D,j}^p\right)^2 \leq \left(S_{DR,j}^{p}\right)^2
\end{equation}
\end{small}
\vspace{-0.9cm}

\subsection{Centralized OPF Problem}
In this section, we define the centralized OPF problems for an unbalanced radial distribution system, that can be solved using a central controller; both nonlinear and linear power flow models have been used for such OPF problems with minimization and maximization objectives.

\subsubsection{Loss Minimization}
First, the active power loss minimization problem has been formulated where the line loss is reduced by generating optimal $q^p_{D,j}$, without violating operational limits. The optimization variable is $X = [P^{pp}_{ij}, Q^{pp}_{ij}, v^p_j, l^{pq}_{ij}, q^p_{D,j}]^T$, $ \forall j\in \N$ \& $\forall \{ij\} \in \E$. The OPF problem \textbf{(C1)} is defined in \eqref{loss_opf_nl} using nonlinear power flow models. Here, $\left(I^{rated}_{ij}\right)^2$ is the thermal limit for the line $\{ij\}\in E$, and $V_{max}$ \& $V_{min}$ denotes the nodal voltage limit.

\vspace{-0.4cm}
\begin{small}
\begin{IEEEeqnarray}{C C}
\IEEEyesnumber\label{loss_opf_nl} \IEEEyessubnumber*
\text{\textbf{(C1)}}  \hspace{0.4cm}  \text{Minimize} \sum_{p \in \phi_j, j:i \rightarrow j}{l^{pp}_{ij}r^{pp}_{ij}}\\
  \text{s.t.}  \hspace{0.3cm}  \text{equation \eqref{nl_pf} \& \eqref{DER_Q} } \label{C1b} \\
  l_{ij}^{p} \leq \left(I^{rated}_{ij}\right)^2  \text{\&} \hspace{0.3cm}
   {V_{min}}^2\leq v_j^p \leq {V_{max}}^2  \label{C1c}
\end{IEEEeqnarray}
\end{small}
\vspace{-0.4cm}

To take advantage of the computational efficacy and scalability, next we define the OPF problem with linear power flow models to minimize the line losses. To further ease the computations,
the nodal voltages are approximated as $1.00$ pu, and thus, the line loss has been approximated by a convex cost function $\sum \left(P^{pp}_{ij}\right)^2+\left(Q^{pp}_{ij}\right)^2$ for all edges $\{ij\}\in \E$. Here for the OPF problem with linear power flow model, the optimization variable is $X = [P^{pp}_{ij}, Q^{pp}_{ij}, v^p_j, q^p_{D,j}]^T$, $ \forall j\in \N$ \& $\forall \{ij\} \in \E$. The OPF problem with linear power flow model, \textbf{(C2)} is defined in \eqref{loss_opf_lin}.

\vspace{-0.4cm}
\begin{small}
\begin{IEEEeqnarray}{C C}
\IEEEyesnumber\label{loss_opf_lin} \IEEEyessubnumber*
\text{\textbf{(C2)}}  \hspace{0.4cm}  \text{Minimize} \sum_{p \in \phi_j, j:i \rightarrow j} {\left(P^{pp}_{ij}\right)^2}+{\left(Q^{pp}_{ij}\right)^2}\\
  \text{s.t.}  \hspace{0.3cm}  \text{equation \eqref{lin_pf} \& \eqref{DER_Q} }  \label{C2b} \\
   {V_{min}}^2\leq v_j^p \leq {V_{max}}^2  \label{C2c}
\end{IEEEeqnarray}
\end{small}
\vspace{-0.5cm}
\subsubsection{DER Maximization}
As a maximization problem, DER maximization objective has been formulated in this paper. Specifically, the network objective is to maximize active power generations, $p_{D,j}^p$, from all the DERs in all of the phases, without exceeding any operational limits. For the DER maximization OPF, the optimization variable is $X = [P^{pp}_{ij}, Q^{pp}_{ij}, v^p_j, l^{pq}_{ij}, p^p_{D,j}]^T$, $ \forall j\in \N$ \& $\forall \{ij\} \in \E$. The OPF problem \textbf{(C3)} is defined in \eqref{der_opf_nl} using the nonlinear power flow models. Similar to the previous OPF problems, $\left(I^{rated}_{ij}\right)^2$ is the thermal limit for the line $\{ij\}\in E$, and $V_{max}$ \& $V_{min}$ denotes the nodal voltage limit.

\vspace{-0.4cm}
\begin{small}
\begin{IEEEeqnarray}{C C}
\IEEEyesnumber\label{der_opf_nl} \IEEEyessubnumber*
\text{\textbf{(C3)}}  \hspace{0.4cm}  \text{Maximize} \sum_{\forall j \in \N_{}}\sum_{p \in \phi_j}{p^{p}_{D,j}}\\
  \text{s.t.}  \hspace{0.3cm}  \text{equation \eqref{nl_pf} \& \eqref{DER_P} }  \label{C3b} \\
  l_{ij}^{p} \leq \left(I^{rated}_{ij}\right)^2  \text{\&} \hspace{0.3cm}
   {V_{min}}^2\leq v_j^p \leq {V_{max}}^2  \label{C3c}
\end{IEEEeqnarray}
\end{small}
\vspace{-0.3cm}

Similar to the problem \textbf{(C2)}, the DER maximization objective problem can be formulated as a linear OPF model with linear approximated power flow model (eqn. \eqref{lin_pf}). Since the objective function for this problem is already a linear cost function, thus the whole OPF problem becomes linear optimization problem. The optimization problem is detailed in \eqref{der_opf_lin} and referred as problem \textbf{(C4)}. The optimization variable for \textbf{(C4)} is $X = [P^{pp}_{ij}, Q^{pp}_{ij}, v^p_j, p^p_{D,j}]^T$, $ \forall j\in \N$ \& $\forall \{ij\} \in \E$.    

\vspace{-0.3cm}
\begin{small}
\begin{IEEEeqnarray}{C C}
\IEEEyesnumber\label{der_opf_lin} \IEEEyessubnumber*
\text{\textbf{(C4)}}  \hspace{0.4cm}  \text{Maximize} \sum_{\forall j \in \N_{}}\sum_{p \in \phi_j}{p^{p}_{D,j}}\\
  \text{s.t.}  \hspace{0.3cm}  \text{equation \eqref{lin_pf} \& \eqref{DER_P} }  \label{C4b} \\
   {V_{min}}^2\leq v_j^p \leq {V_{max}}^2  \label{C4c}
\end{IEEEeqnarray}
\end{small}
\vspace{-0.7cm}

\section{Distributed Algorithm}
The optimization problem for radial power distribution networks-- composed of several connected areas, are naturally decomposable into multiple sub-problems defined for the areas. In this section, first we detail the decomposition approach for the OPF problems of three-phase unbalanced systems, and then the local sub-problems are defined. Lastly, we briefly discuss the developed cyber-physical co-simulation platform, where both the physical and the communication layer are simulated{\color{black} to study the impact of the developed distributed OPF algorithm on a real life distribution network.}

\vspace{-0.3cm}
\subsection{Decomposition Method \& Network Approximations}
{\color{black}Let the distribution system be composed of $N$ connected areas/microgrids, and $\A = \{A_1,~A_2, …,~A_{N}\}$ be the set of all areas or microgrids. Any area $A_m \in \A$ is defined as the directed radial graph $A_m = \G(\N_m, \E_m)$; $\N_m$ and $\E_m$ defines the set of nodes and lines in the system. The set of central optimization variables is defined as $X = \bigcup_{m=1}^{N} X_m$, where $X_m $ represents the set of all local optimization variables for Area $A_m$. Similarly $\mathcal{S} = \bigcup_{m=1}^{N}\mathcal{S}_m$ represent the set of constraints for overall problem, and $\mathcal{S}_m$ represents the local constraints. If central objective $f(X)$ is a decomposable cost function, then the problem can be decomposed into several local sub-problems and written as \eqref{eqProposed_obj}. Here, $f_m$ is defined as the local objective function for Area $A_m$. The complicating variables, $y_{{m}^'}$ are shared by the neighboring areas of $A_m$, and are kept constant for solving the local optimization problem in this primal decomposition-based method.

\vspace{-0.3cm}
\begin{equation}\label{eqProposed_obj}
\small
\min_{X \in \mathcal{S}}{\hspace{0.1cm} f(X)} = {\hspace{0.1cm}} {\sum_{m=1}^N \min_{X_m \in \mathcal{S}_m} f_m(X_m,y_{{m}^'})}
\end{equation}
\vspace{-0.3cm}



As the overall system topology was radial to begin with, the decomposed areas of the networks are also connected radially and have unique upstream \& downstream sections. This specific structure of the system helps to identify the unique parent area and the child areas for any area $A_m$, which in turns associates the complicating/shared variables, $y_{{m}^'}$ for the local sub-problems -- exchanged among the neighboring computing agents to solve the overall problem. For the proposed decomposition approach, the complicating variables are the voltages and the power flows at the shared bus. Let for any Area $A_m$, bus $0$ is shared with its parent/Upstream area (UA) and node $k$ ($\{jk\} \in \E_m$) is shared with its child/Downstream Area (DA). Then the complicating variable, $y_{{m}^'} = [y^{ua},y^{da}]$, where $y^{ua} = [v_{0_{ua}}]$ , and $y^{da} = [P_{jk_{da}}, Q_{jk_{da}}]$. Here, each area solves related local sub-problems in parallel by assuming the constant complicating variable, i.e., a fixed voltage at the shared bus with the unique UA and constant loads at the shared buses with DAs. In terms of power system's physics, this approximates the whole upstream and downstream network segments of an area for radial power distribution systems.}

After solving local sub-problems, the respective complicating variables, i.e., the total power requirements in that area and the shared (with DA) bus voltages area are shared with UA and DAs, respectively. Then, the sub-problems are solved again with updated shared variables from the neighbors. When a consensus is attained at the shared boundaries, then the decision variables are dispatched within the area. The consensus at the boundary can be achieved using Fixed Point Iteration methods as described in \cite{sadnan2021online}.

\begin{figure*}[t]
    \centering
    \includegraphics[width=0.85\textwidth]{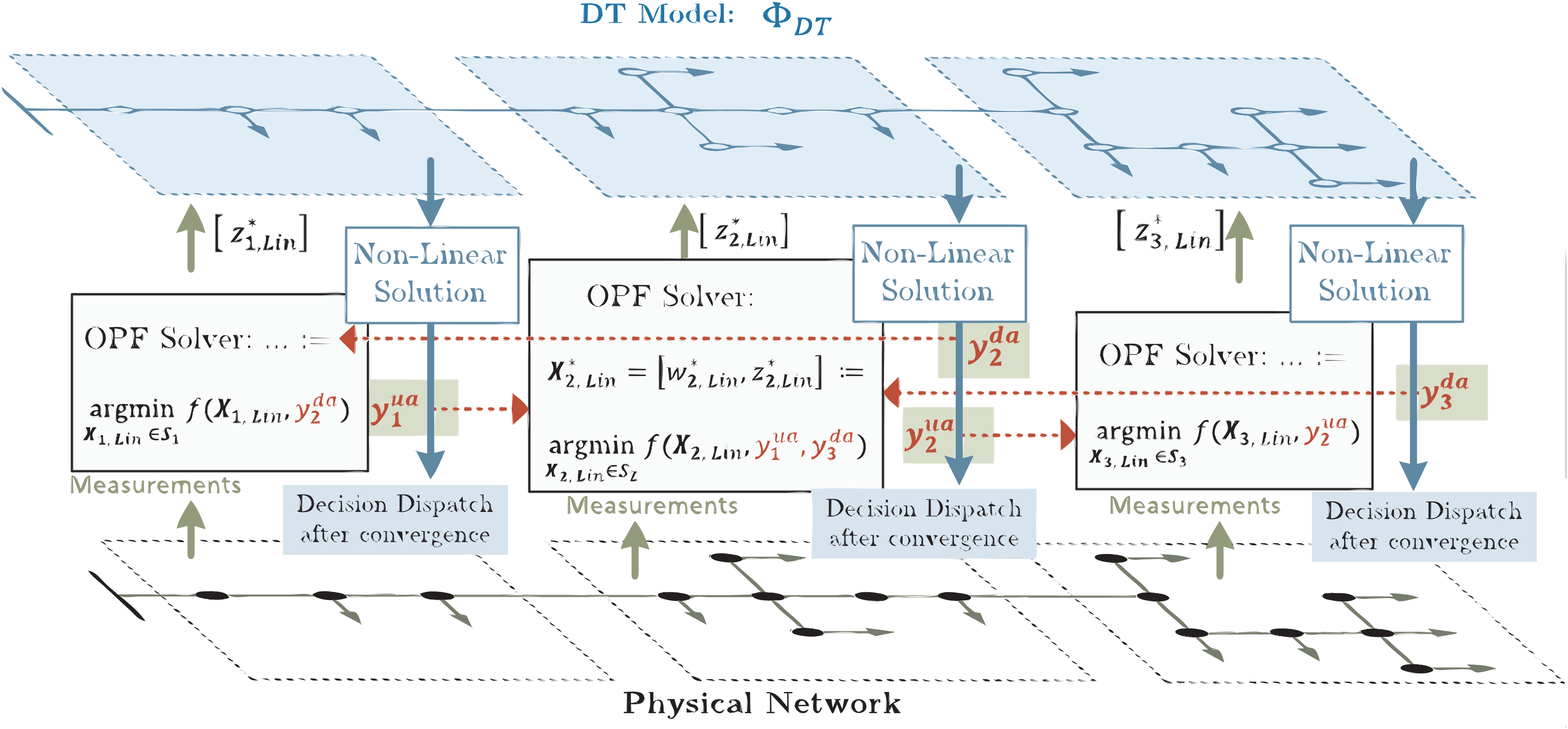}
        \vspace{-0.2cm}
    \caption{Use of Digital Twin Model in Distributed Optimization for Power System}
    \label{DTC}
\end{figure*}

\subsection{Local Distributed OPF Model}
In this section, the decomposed local sub-problems at $n^{th}$ macro-iteration, \textbf{(D1)-(D4)}, of the corresponding central OPFs, \textbf{(C1)-(C4)} are defined. For any Area $A_m$, let bus $0$ is shared with its UA and node $k$ ($\{jk\} \in \E_m$) is shared with its DAs. Also, the subscript of a node represents the area, that has solved the variable; e.g., $v_{0_m}$, $p_{k_m}$ are solved by area $A_m$, where $v_{0_{ua}}$, $P_{jk_{da}}$ are solved by UA and DA, respectively. In the proposed model, each local computing agents solves the corresponding distributed sub-problems, and over macro-iterations, convergence is achieved.

\subsubsection{Loss Minimization} The OPF problem \textbf{(C1)-(C2)} represent the central formulation of the loss minimization problem using nonlinear and approximated models, respectively, where the active power line losses are minimized by controlling the reactive power generations from the DERs. The corresponding decomposed sub-problems for Area $A_m$ , $\forall \{ij\}\in \E_m$ and $\forall j \in \N_m$  are defined in \eqref{D1}-\eqref{D2}, respectively. Constraint \eqref{eqvD1} and \eqref{eqvD2}  represents the approximated UA and DA with the help of complicating/shared variables, as described in the previous section. 

\vspace{-0.3cm}
\begin{small}
\begin{IEEEeqnarray}{C C}
\IEEEyesnumber\label{D1} \IEEEyessubnumber*
\text{\textbf{(D1)}}\hspace{0.4cm} \min \hspace{0.2cm} f_m(X_m^{(n)}) =\sum_{p \in \phi_j, j:i \rightarrow j}{l^{pp}_{ij}r^{pp}_{ij}} \\
\text{s.t. equation \eqref{C1b} \& \eqref{C1c} } \label{conD1}\\
\hspace{-2pt}v_{0_m}^{p\hspace{1pt} (n) } = v_{0_{ua}}^{p\hspace{1pt}(n-1) }\label{eqvD1}; \hspace{0.1cm}
p_{k_m}^{p \hspace{1pt}(n)} = P_{jk_{da}}^{pp \hspace{1pt}(n-1)}; \hspace{0.1cm}
q_{k_m}^{p \hspace{1pt}(n)} = Q_{jk_{da}}^{pp \hspace{1pt}(n-1)}
\end{IEEEeqnarray}
\end{small}
\vspace{-0.3cm}

\begin{small}
\begin{IEEEeqnarray}{C C}
\IEEEyesnumber\label{D2} \IEEEyessubnumber*
\text{\textbf{(D2)}}\hspace{0.4cm} \min \hspace{0.2cm} f_m(X_m^{(n)}) =\sum_{p \in \phi_j, j:i \rightarrow j} {\left(P^{pp}_{ij}\right)^2}+{\left(Q^{pp}_{ij}\right)^2}\\
\text{s.t. equation \eqref{C2b} \& \eqref{C2c} } \label{conD2}\\
\hspace{-2pt}v_{0_m}^{p\hspace{1pt} (n) } = v_{0_{ua}}^{p\hspace{1pt}(n-1) }\label{eqvD2}; \hspace{0.1cm}
p_{k_m}^{p \hspace{1pt}(n)} = P_{jk_{da}}^{pp \hspace{1pt}(n-1)}; \hspace{0.1cm}
q_{k_m}^{p \hspace{1pt}(n)} = Q_{jk_{da}}^{pp \hspace{1pt}(n-1)}
\end{IEEEeqnarray}
\end{small}

\subsubsection{DER Maximization} The OPF problem \textbf{(C3)-(C4)} represent the central OPF formulation of the DER generation maximization problem using nonlinear and approximated models, respectively, where the real power generations for all the phases are maximized within the physical limits and without violating the operational constraints. Similar to loss minimization objective, the corresponding decomposed sub-problems for Area $A_m$ are defined in \eqref{D3}-\eqref{D4}, respectively. Similar to loss minimization problem, constraint \eqref{eqvD3} and \eqref{eqvD4}  represents the approximated UA and DA with the help of complicating/shared variables. 

\vspace{-0.3cm}
\begin{small}
\begin{IEEEeqnarray}{C C}
\IEEEyesnumber\label{D3} \IEEEyessubnumber*
\text{\textbf{(D3)}}\hspace{0.4cm} \max \hspace{0.2cm} f_m(X_m^{(n)}) =\sum_{\forall j \in \N_{m}}\sum_{p \in \phi_j}{p^{p}_{D,j}}\\
\text{s.t. equation \eqref{C3b} \& \eqref{C3c} } \label{conD3}\\
\hspace{-2pt}v_{0_m}^{p\hspace{1pt} (n) } = v_{0_{ua}}^{p\hspace{1pt}(n-1) }\label{eqvD3}; \hspace{0.1cm}
p_{k_m}^{p \hspace{1pt}(n)} = P_{jk_{da}}^{pp \hspace{1pt}(n-1)}; \hspace{0.1cm}
q_{k_m}^{p \hspace{1pt}(n)} = Q_{jk_{da}}^{pp \hspace{1pt}(n-1)}
\end{IEEEeqnarray}
\end{small}
\vspace{-0.9cm}

\begin{small}
\begin{IEEEeqnarray}{C C}
\IEEEyesnumber\label{D4} \IEEEyessubnumber*
\text{\textbf{(D4)}}\hspace{0.4cm} \max \hspace{0.2cm} f_m(X_m^{(n)}) =\sum_{\forall j \in \N_{m}}\sum_{p \in \phi_j}{p^{p}_{D,j}}\\
\text{s.t. equation \eqref{C4b} \& \eqref{C4c} } \label{conD4}\\
\hspace{-2pt}v_{0_m}^{p\hspace{1pt} (n) } = v_{0_{ua}}^{p\hspace{1pt}(n-1) }\label{eqvD4}; \hspace{0.1cm}
p_{k_m}^{p \hspace{1pt}(n)} = P_{jk_{da}}^{pp \hspace{1pt}(n-1)}; \hspace{0.1cm}
q_{k_m}^{p \hspace{1pt}(n)} = Q_{jk_{da}}^{pp \hspace{1pt}(n-1)}
\end{IEEEeqnarray}
\end{small}
\vspace{-0.3cm}

\vspace{-0.6cm}
\subsection{Simulation Within Optimization: Use of Digital Twin}
To reduce the compute burdens to solve the NLP optimization problem, a method utilizing simulation within optimization has been proposed by incorporating the Digital Twin (DT) of the system. DTs are the detailed virtual model of the real-world system/network, linked to their physical counterpart by the continuous, real-time information flow. Specific to the power distribution systems, DTs are the software-based abstractions of complex physical networks of power distribution systems which are connected via a communication link to the actual physical system; real-time analysis can be conducted on these virtual models/DTs for decision support purposes. By using the DT models for solving optimization problems, we can examine the optimal solutions ahead of time, and without enacting the decision variables directly in the real systems, the feasibility of the systems can be ensured -- preventing the potential unwanted events for the network. This unfolds the opportunity to use approximated linear power flow models for solving OPFs -- defined in equation \eqref{lin_pf}, instead of using computationally complex nonlinear network models as described by equation \eqref{nl_pf}. {\color{black} Upon implementing the optimal decision variables in the DT models (e.g., optimal DER generations -- obtained from the approximated models), the corresponding nonlinear solutions of the state variables of the system, such as nodal voltages, branch power flows etc., of the system can be attained.

Mathematically, the optimization variable $X$ for the OPF problems is composed of state variables, $w$, and the optimization decision variables, $z$, i.e., $X = [w, z]^T$. For example, in the \textbf{(C1)} OPF problem, $w = [P^{pp}_{ij}, Q^{pp}_{ij}, v^p_j, l^{pq}_{ij}]$, and $z = q^p_{D,j}$ $ \forall j\in \N$ \& $\forall \{ij\} \in \E$. Similarly for approximated \textbf{(C4)} OPF problem, $w = [P^{pp}_{ij}, Q^{pp}_{ij}, v^p_j]$, and $z = p^p_{D,j}$ $ \forall j\in \N$ \& $\forall \{ij\} \in \E$. We use DT models for the simulation within optimization as such:
after solving the approximated OPF problem (equation \eqref{DT1}), the optimal decision variables, $z_{Lin}^*$ are implemented in the DT model -- defined by the function $\Phi_{DT}$, and then the nonlinear solutions are extracted. Thus, using \eqref{DT2}, the corresponding nonlinear state variables of the system, $w_{Nl}^*$ are obtained upon effectuating the linear optimizer $z_{Lin}^*$ in the system; the optimizer $X^*$ attained from this simulation within optimization approach is shown in equation \eqref{x_star}. Any nonlinear power-flow model of the system can be used to represent the functional form for the DT.}

\vspace{-0.3cm}
\begin{small}
\begin{equation}\label{DT1}
X_{Lin}^*=[w_{Lin}^*,z_{Lin}^*] := \underset{X_{Lin} \in \mathcal{S}}{\text{argmin}} {\hspace{0.1cm} f(X_{Lin})}
\end{equation}
\vspace{-0.3cm}
\begin{equation}\label{DT2}
w_{Nl}^* = \Phi_{DT} (z_{Lin}^*)
\end{equation}
\vspace{-0.5cm}
\begin{equation}\label{x_star}
X^* = [w^*_{Nl}, z_{Lin}^*]^T
\end{equation}
\end{small}
\vspace{-0.4cm}

Similarly, in the distributed decision-making process, the use of DTs for solving the OPF problems helps to build a routine that can leverage the computational speed and scalability of the approximated optimization problems with linear power flow models. Each distributed agent has the DT models ($\Phi_{DT}$) of the corresponding local networks. The agents first solve the local approximated sub-problems (similar to \eqref{DT1}), and then extract the nonlinear solutions by using \eqref{DT2} for the local system. During the macro-iteration, i.e., communication among these distributed agents, appropriate variables from nonlinear solution, $w_{Nl}^*$ are exchanged, instead of using the solutions from linear models, i.e., $w_{Lin}^*$.
Specifically, alternate to the direct exchange of linear solutions, the nonlinear solutions from DT simulations are shared among the distributed agents. This effectively helps to ensure the feasibility of the optimal solution of the overall problem at the end of the macro-iterations, even though the computationally simple and scalable approximated models are being solved. {\color{black} The block diagram of the proposed distributed algorithms with simulation within optimization using DT models is shown in Fig. \ref{DTC}}

\subsection{Co-simulation Platform}
\begin{figure}[t]
    \centering
    \includegraphics[width=0.35\textwidth]{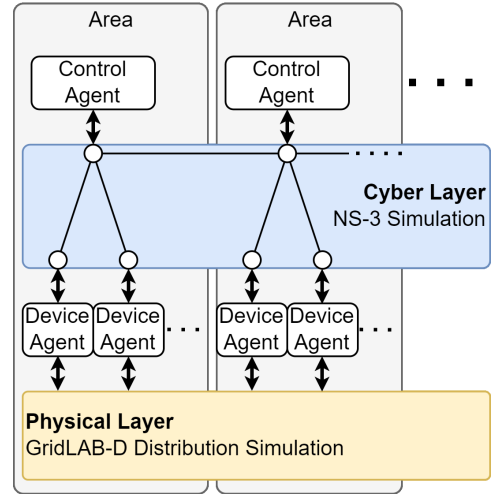}
    \caption{Co-Simulation Structure for Cyber-Physical System}
    \label{cosim}
\end{figure}

A realistic evaluation of distributed algorithms require modeling the associated communication system. To this end, we developed a multi-agent cyber-physical co-simulation package using HELICS (Heirarchical Engine for Large-scale Infrastructure Co-Simulation), a generic co-simulation platform  \cite{palmintier2017design}. Our cyber-physical co-simulation package is composed of the following software: Python 3, GridLAB-D \cite{Chassin:2008fq}, NS-3 \cite{ns3}, which are used to simulate the computing agents, the power distribution networks, the communication networks, respectively. HELICS \cite{palmintier2017design} is then employed to synchronize the simulations and facilitates message passing among the three simulation software (see Fig.~\ref{cosim}). Each area has a controller agent shown at the top which has a single connection, via HELICS, to the communication network simulated in NS-3. In this way, the controllers are able to communicate with each other and with the device agents in the local area. The device agents  each have a connection, via HELICS, to the physical device simulated in GridLAB-D and to the communication network. The purpose of the device agent is twofold. It formats the raw data it receives from GridLAB-D so the control agent can understand it and it parses commands received from the control agent to the GridLAB-D. The control agents are capable of running their own GridLAB-D simulations using local area models as digital twins (DT). Each area employs the respective DT to obtain the actual power flow variables the system observes upon implementing the approximated OPF control variables.

\vspace{-0.2cm}
\section{Results and Discussions}
\vspace{-0.2cm}
In this section, a detailed evaluation of the proposed D-OPF algorithm for a 3-phase unbalanced distribution system is presented. First, we have validated our distributed OPF approach for nonlinear optimization problems by comparing the solution with the central formulation. Then, we have simulated and verified the simulation within optimization approach that has been proposed in this paper with cyber-physical co-simulation platform and uses approximated OPF problems. Lastly, a time-series simulation has been performed to showcase the effect of network optimization using the proposed method, and different stressed communication network topologies have been tested to demonstrate the efficacy and robustness of the proposed distributed OPF algorithm.
\begin{figure}[t]
    \centering
    \includegraphics[width=0.43\textwidth]{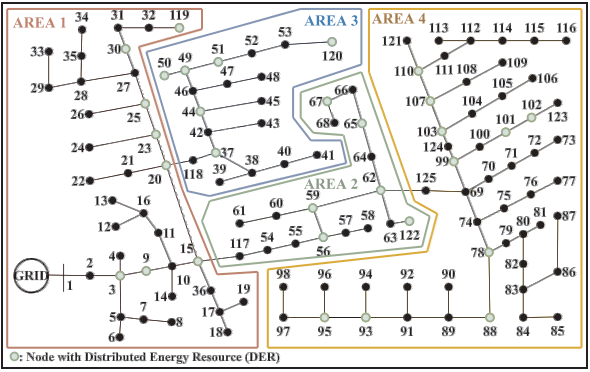}
        \vspace{-0.2cm}
    \caption{Sectioned IEEE 123 Bus System}
    \label{123bus}
\end{figure}

\subsection{Simulated System}
To test and validate our proposed method, we have simulated two network objective problems with three-phase unbalanced IEEE-123 bus test system that has been assumed to be composed of 4 areas. Three different scenarios of DERs are simulated in this paper to test the proposed method -- (i) 10 three-phase DERs (at node 15, 23, 30, 37, 49, 50, 51, 67, 78, 107) with capacity of 60 kVA per phase, (ii) 30 three-phase DERs with capacity of 48 kVA per phase, and (iii) 30 three-phase DERs with capacity of 60 kVA per phase (Fig. \ref{123bus}). Please note, scenario (ii) is simulated for loss minimization problem to provide a more challenging problem for the algorithm in terms of required communication properties. Further, for the time series simulation, the loads are also varied randomly between 78\% to 98\% of the nominal values for 60 minutes. In addition to that, as a digital twin function, $\Phi_{DT}$, GridLab-D power flow simulator has been used in this study.

\begin{figure*}[t]
\centering
\begin{subfigure}{.32\textwidth}
  \centering  \includegraphics[width=1.0\linewidth]{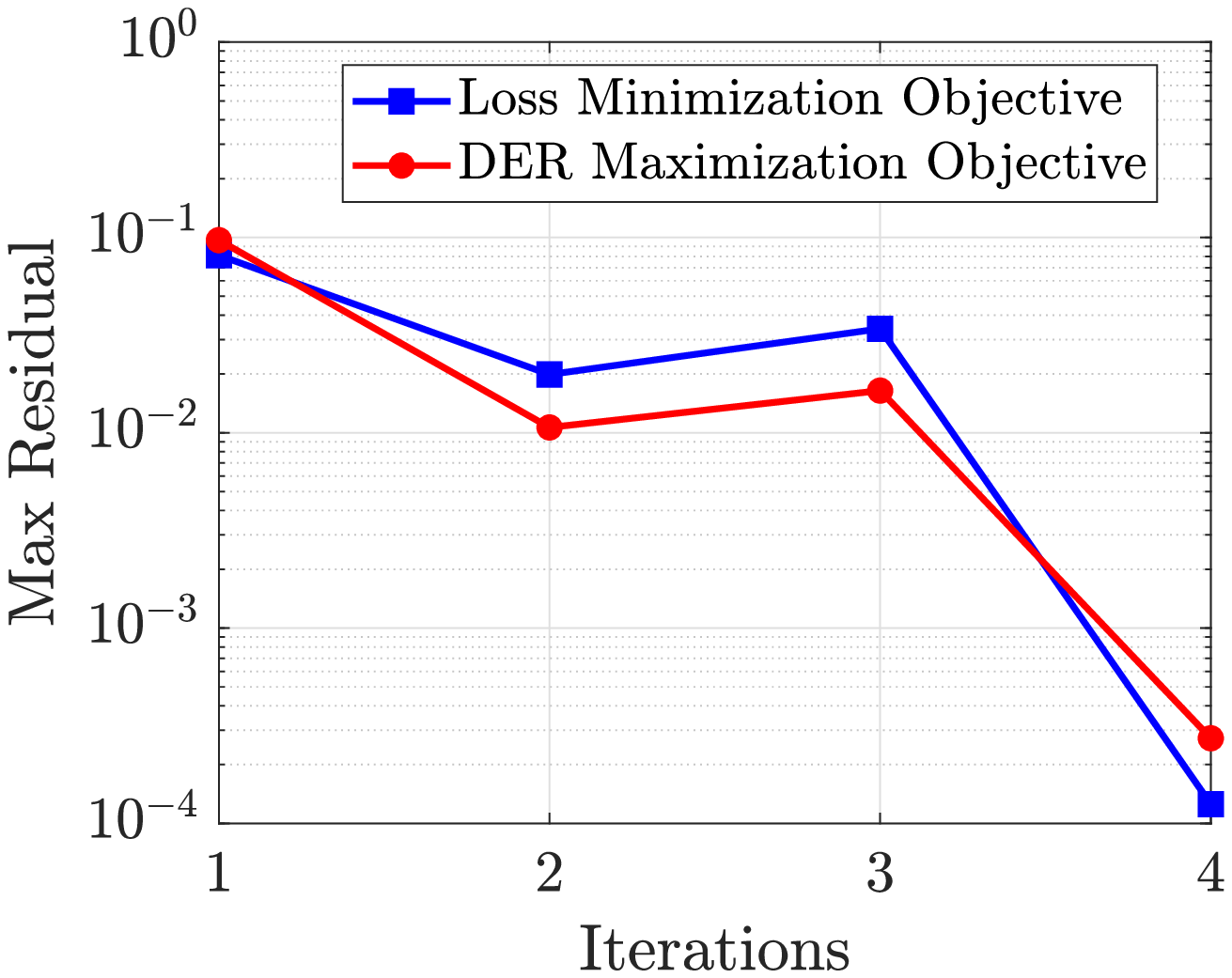}
  \vspace{-0.6cm}
  \caption{Convergence at Boundary}
  \label{residual_cmp}
\end{subfigure}
\begin{subfigure}{.32\textwidth}
  \centering
  \includegraphics[width=1.0\linewidth]{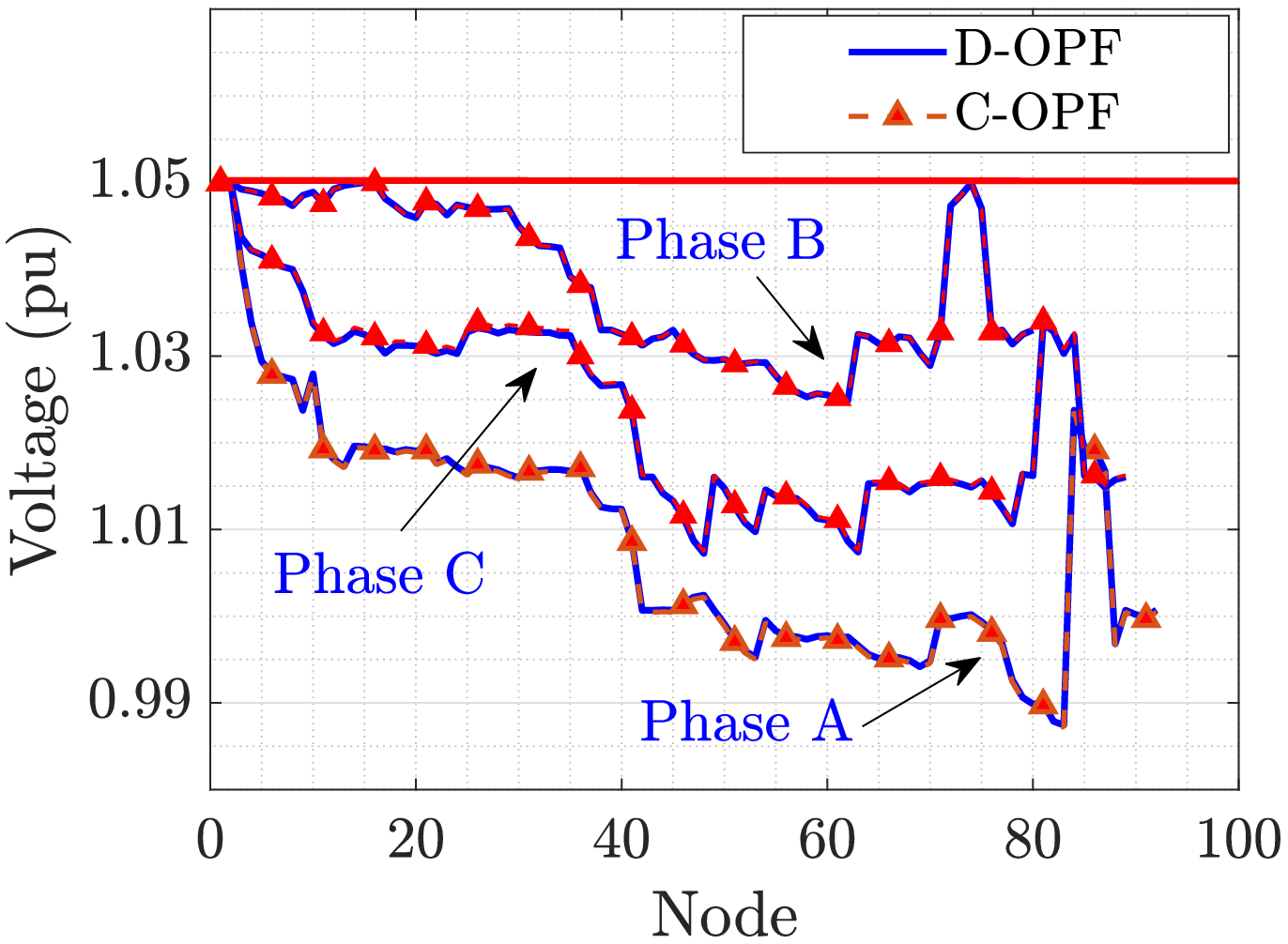}
  \vspace{-0.6cm}
  \caption{Node Voltages: Loss Minimization}
  \label{volt_lm}
\end{subfigure}
\begin{subfigure}{.32\textwidth}
  \centering
  \includegraphics[width=1.0\linewidth]{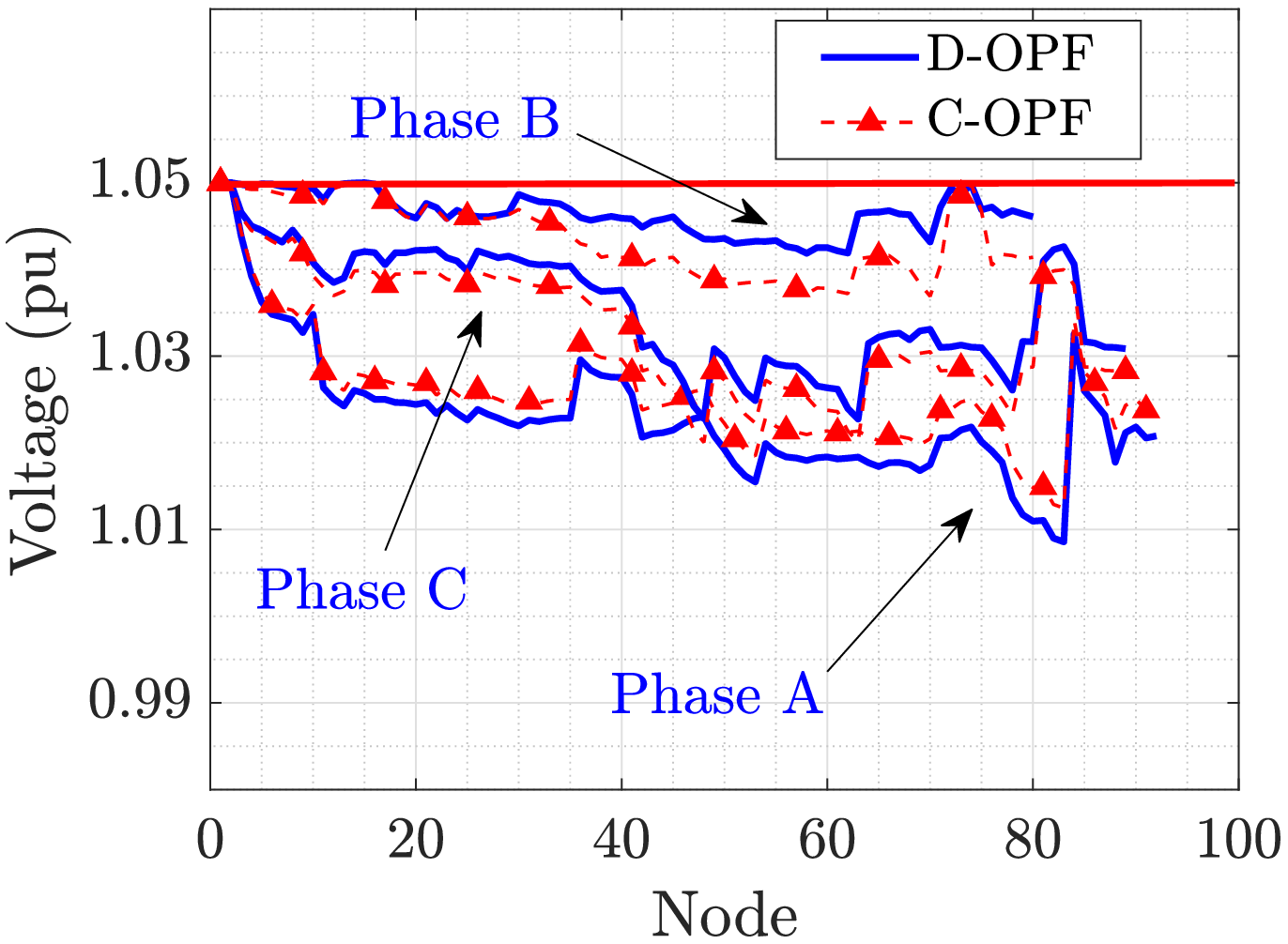}
  \vspace{-0.6cm}
  \caption{Node Voltages: DER Maximization}
  \label{volt_pvc}
\end{subfigure}
\vspace{-0.2cm}
\caption{D-OPF Results for Loss Minimization and DER Maximization Objectives}
\label{D_nlp}
\end{figure*}

\subsection{Validation of Decomposition Method}
\begin{figure}[h]
\centering
\begin{subfigure}{.24\textwidth}
  \centering
  \includegraphics[width=1.1\linewidth]{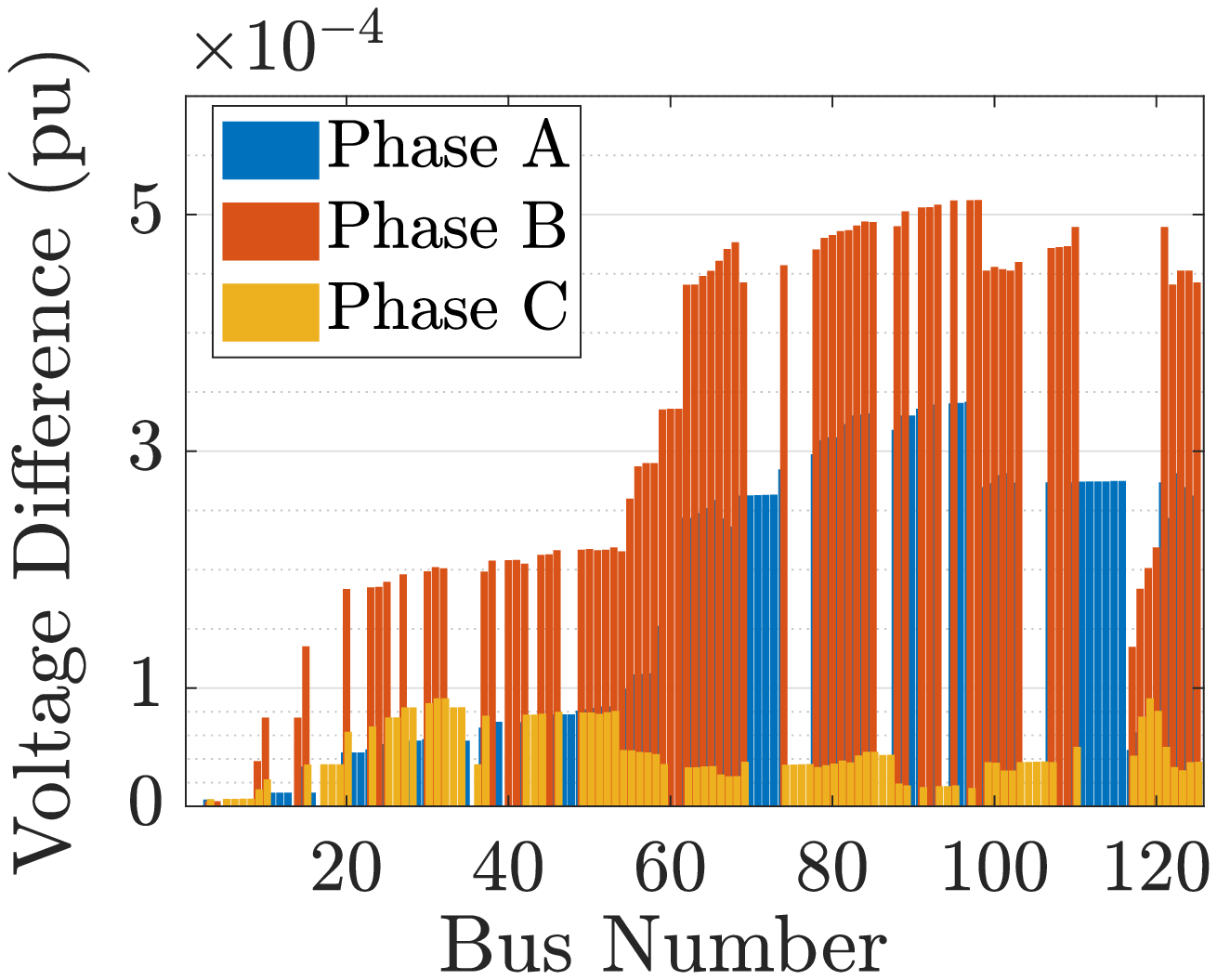}
  \caption{Loss Minimization}
  \label{val_lm}
\end{subfigure}
\begin{subfigure}{.24\textwidth}
  \centering
  \includegraphics[width=1.1\linewidth]{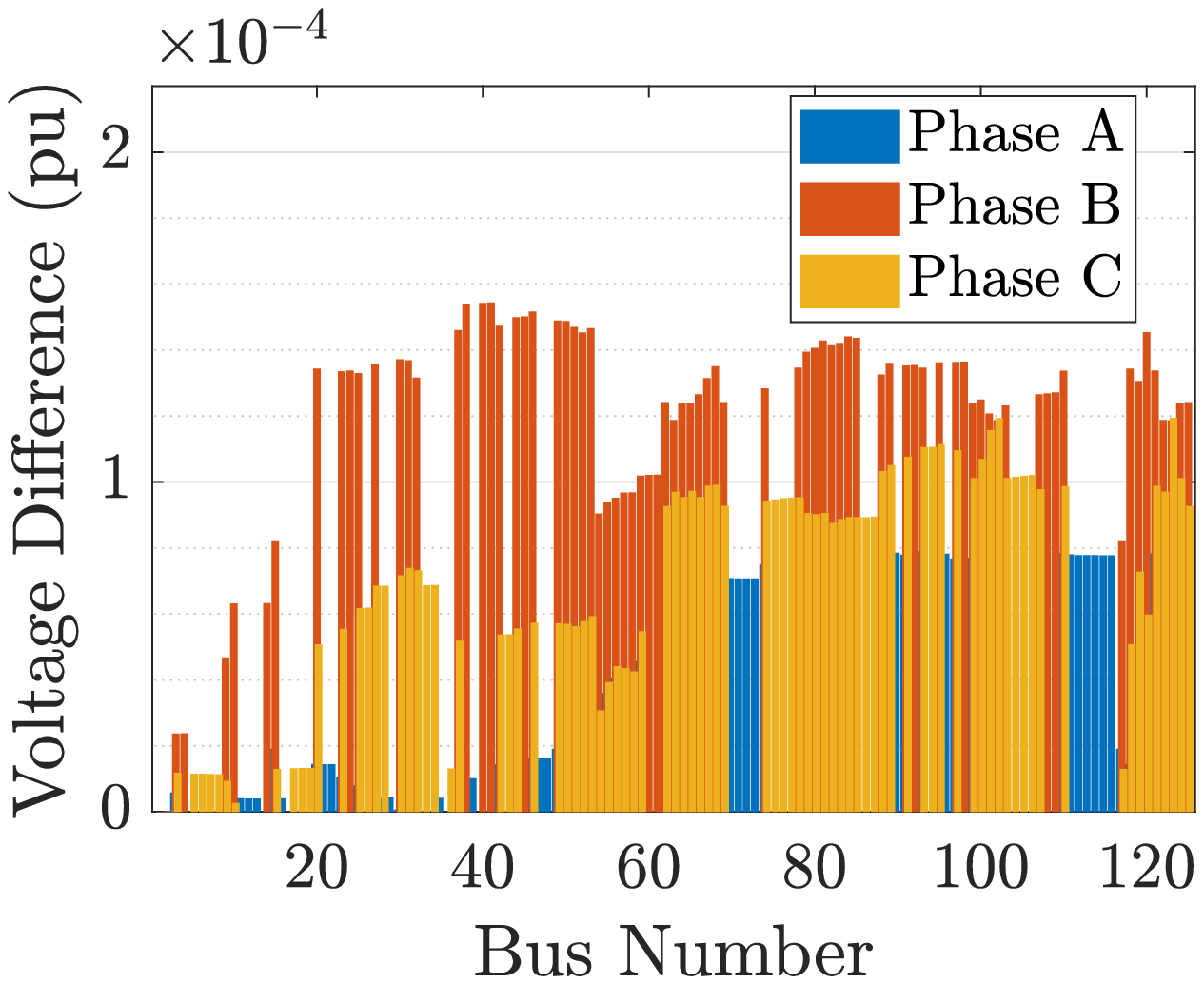}
  \caption{DER Maximization}
  \label{val_pvc}
\end{subfigure}
\caption{Validation for Decomposition Approach}
\label{val_nlp}
\end{figure}

To validate the proposed D-OPF method for unbalanced power distribution systems, we have simulated two optimization problems,\textbf{ (D1)} \& \textbf{(D3)} for DER scenarios (i) \& (iii), respectively, and compared with the solutions of \textbf{(C1)} \& \textbf{(C3)} (see Table \ref{NL_val}). It is evident that the optimal solutions of D-OPF match with the corresponding Central OPF (C-OPF) solutions for both objectives. For example, the objective value for the DER maximization problem, i.e., total DER generation is 5.08 MW \& 5.18 MW, when solved using D-OPF \& C-OPF method, respectively. However, D-OPF takes significantly less time to solve, compared to the C-OPF method -- takes almost 10 times or more for some cases. From Fig. \ref{residual_cmp}, it can be seen that it only takes 4 macro-iterations to reach the global solution. In addition to that, the nodal voltages of D-OPF and C-OPF solutions are compared (Fig. \ref{volt_lm}-\ref{volt_pvc}). Later the nodal voltages of D-OPFs are compared with the GridLab-D/OpenDSS model -- upon enacting the decision variables from D-OPF, and the voltage difference is in the order of $10^{-4}$ or less (Fig. \ref{val_nlp}). These comparisons validate the feasibility and the optimality of the proposed D-OPF solutions for both minimization and maximization objectives.

\begin{small}
\begin{table}
    \centering
    \caption{Validation of D-OPF method}
    \setlength{\tabcolsep}{3pt}
    \begin{tabular}{c  c c  c c }
    \toprule
    \multirow{2}{*}{\textbf{OPF Problem}}  & \multicolumn{2}{c}{\textbf{Objective}} & \multicolumn{2}{c}{\textbf{Time (s)}} \\
    & D-OPF & C-OPF & D-OPF & C-OPF \\
    \midrule
    Loss Min.: Scenario (i)  & 43.976 kW & 43.975 kW & 20 & 520 \\
    DER Max.: Scenario (iii)    & 5.08 MW & 5.18 MW & 20 & 200 \\ 
    \bottomrule
    \end{tabular}
    \label{NL_val}
\end{table}
\end{small}

\subsection{Co-simulation \& Use of DT}
\vspace{-0.1cm}
Next, to showcase the robustness and the practicability of the proposed algorithm, the D-OPF method is implemented on a cyber-physical co-simulation platform. For this case, distributed problems \textbf{(D2) \& (D4)} are solved to leverage the simplicity of the linear network model for edge computing use-cases to solve the OPFs. In this paper, the DTs of the local areas are modeled using GridLab-D power flow simulator. For this paper, it is assumed that the distributed agents/local controllers communicate with neighbor every 2-seconds. And also, after the convergence is attained, it takes around $0.2$ seconds to dispatch the decision variables in the system.

\begin{small}
\begin{table}
    \centering
    \vspace{-0.15cm}
    \caption{Loss minimization Result Comparison}
    \vspace{-0.1cm}
    \label{loss_min10}
    \setlength{\tabcolsep}{3pt}
    \begin{tabular}{c c c c c}
    \toprule
    
    {\textbf{Method}}  & \multicolumn{2}{c}{\textbf{Loss (kW)}}  & \multicolumn{2}{c}{\textbf{Communication-round}}\\
    DER Scenario & Scenario (i) & Scenario (ii) &Scenario (i) &Scenario (ii)\\
    \midrule
    No OPF                     & 63.132 & 53.338 & 0 & 0 \\
    Linear C-OPF               & 44.875 & 26.508 & 0 & 0\\
    Linear D-OPF               & 44.959 & 26.546 & 5 & 5 \\ 
    Linear D-OPF+DT            & 44.959 & 27.238  & 5 & 5 \\
    Nonlinear C-OPF              & 43.975 & 27.160 & 0 & 0\\
    
    \bottomrule
    \end{tabular}
    \vspace{-0.6cm}
\end{table}
\end{small}

\subsubsection{Validation}
Now we validate the co-simulation platform using the linear approximate models by comparing the resulting objective values. Table \ref{loss_min10} and Table \ref{der_max}, show the comparisons between D-OPF and C-OPF results for different cases. Table~\ref{loss_min10} shows the loss minimization case for DER scenarios (i) and (ii), where real power generations are fixed at $50$ and $20$ kW per phase, respectively. For both DER scenarios, line losses matches with the nonlinear solutions. For example, in scenario (i), the active power loss is $44.959$ kW for linear model with the DT model, and the nonlinear solution is $43.975$ kW. However, for scenario (ii), it can be seen that the use of DT model has be potential to get closer solutions to the nonlinear solution. In Table \ref{loss_min10}, the line loss for scenario (ii) is $26.546$ kW when DT models are not being used, however, if the DT models are used, the objective value is $27.23$ kW -- that is closer to the nonlinear solution $27.16$ kW. The use of DT with approximated model does not necessarily improve the results, but it guarantees the feasibility of the local solutions, before exchanging the boundary variables with the neighbors. Also, it takes around 5 communication-rounds among the agents to reach the consensus at the boundary. Similarly, Table \ref{der_max} validates the algorithm on the co-simulation platform for DER maximization problem that uses DER scenario (iii). When no OPF is performed, there were voltage limit violation in the system. However, if operational limits are considered, the optimal generation is $5.135$, $5.163$, and $5.18$ MW for linear D-OPF, linear D-OPF+DT, and nonlinear C-OPF, respectively. Thus the use of the DT model for DER maximization problem also brings the solution closer to the actual nonlinear solutions. Also, using DT model reduces the number of communication rounds from 7 to 5. Thus, Table \ref{loss_min10} and Table \ref{der_max} validate the co-simulation platform that uses computationally simple linear approximated OPF model with the DT model of the system.

\begin{small}
\begin{table}
    \vspace{-0.1cm}
    \centering
    \caption{DER Maximization Results Comparison}
    \vspace{-0.1cm}
    \label{der_max}
    \setlength{\tabcolsep}{3pt}
    \begin{tabular}{c  c  c}
    \toprule
    \textbf{Method}  & \textbf{Generation (MW)} & \textbf{Communication-round}\\
    DER Scenario & Scenario (iii) & Scenario (iii)\\
    \midrule
    No OPF                     & 5.400 & 0\\
    Linear C-OPF               & 5.135& 0\\
    Linear D-OPF               & 5.135 & 7\\ 
    Linear D-OPF+ DT           & 5.163 & 5\\ 
    Nonlinear C-OPF              & 5.180 & 0\\
    \bottomrule
    \end{tabular}
    \vspace{-0.7cm}
\end{table}
\end{small}

\subsubsection{Time-series Simulation}
\begin{figure}[t]
\centering
\begin{subfigure}{.24\textwidth}
  \centering
  \includegraphics[width=1.1\linewidth]{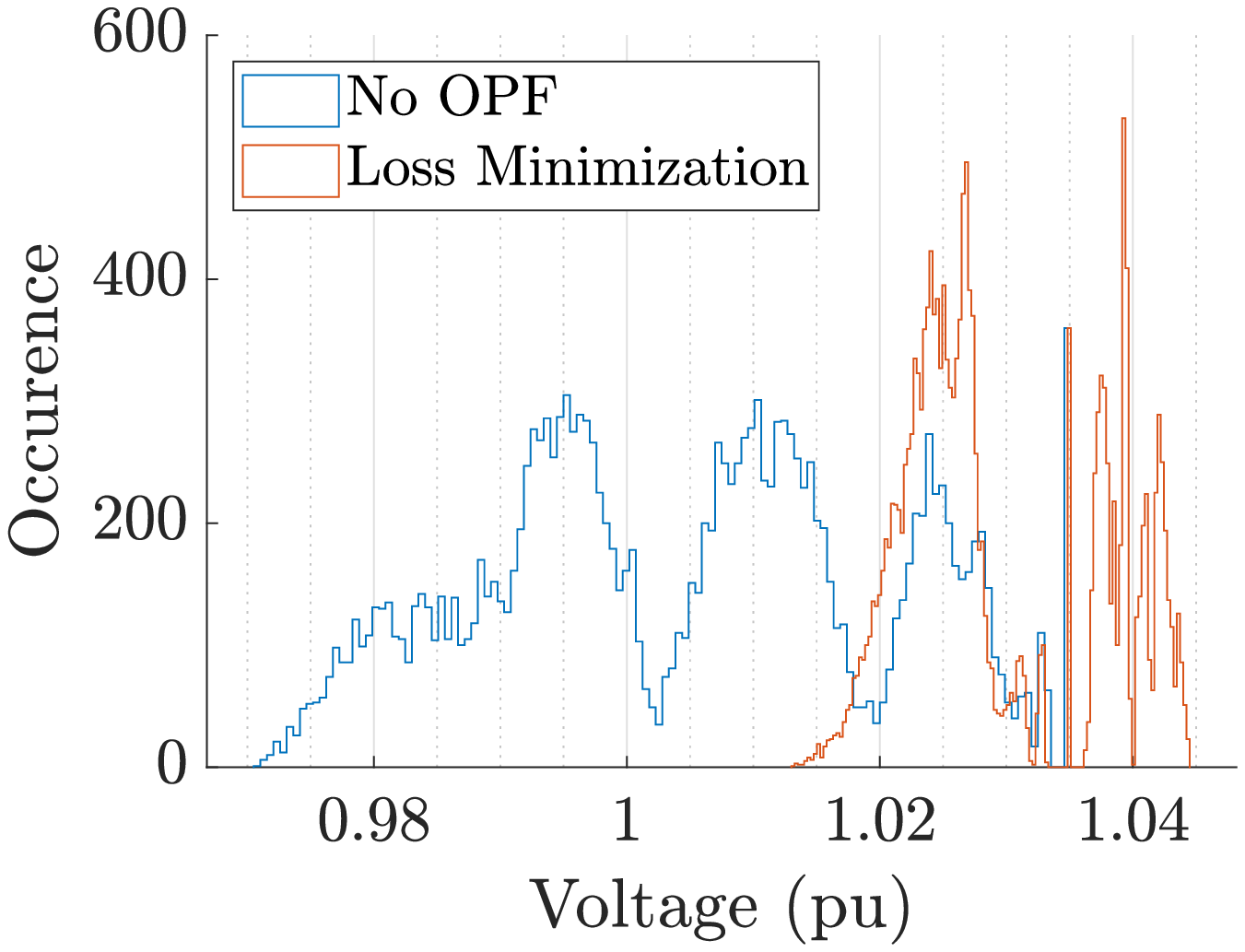}
  \vspace{-0.6cm}
  \caption{Loss Minimization}
  \label{dist_lm}
\end{subfigure}
\begin{subfigure}{.24\textwidth}
  \centering
  \includegraphics[width=1.1\linewidth]{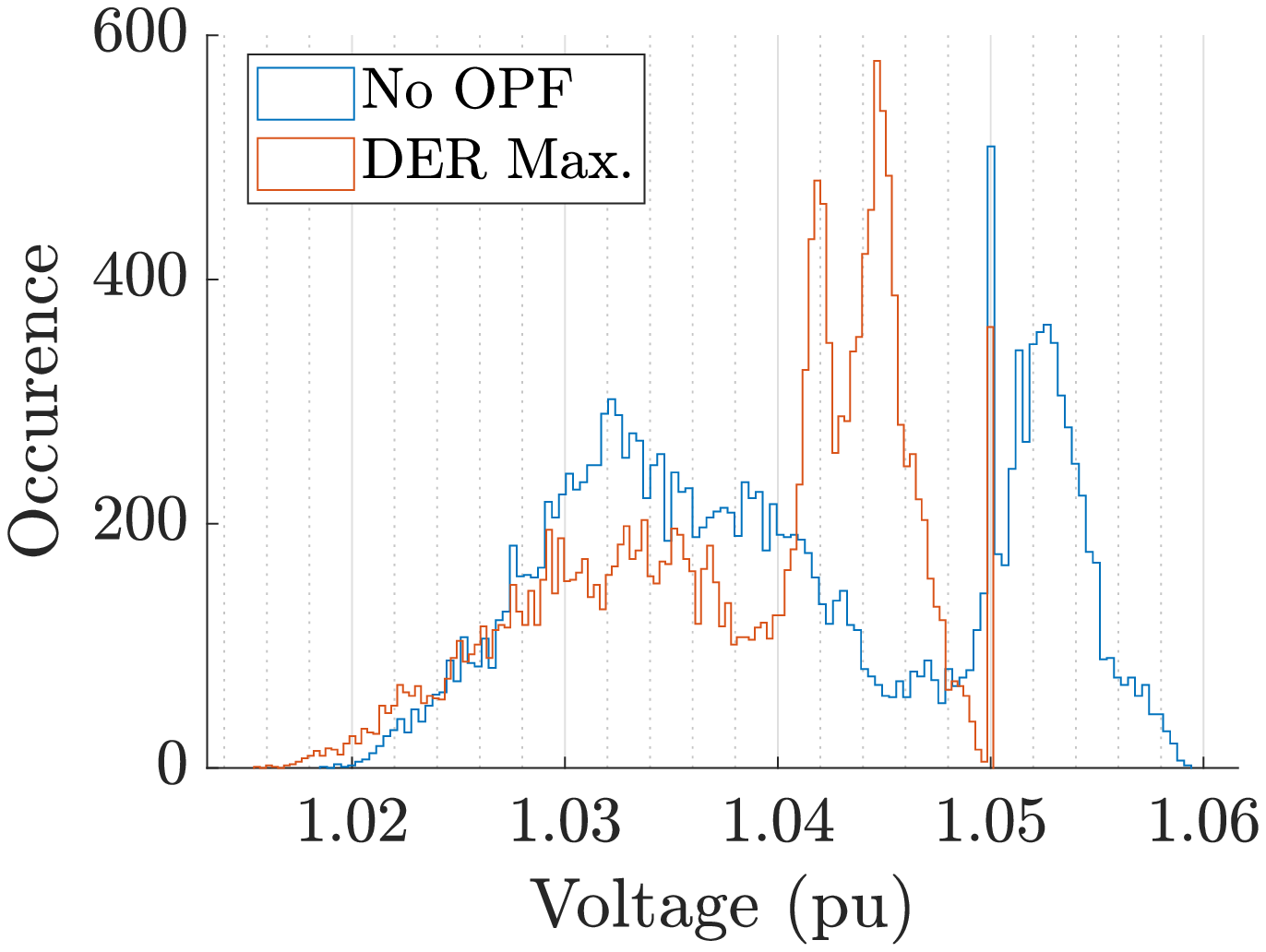}
  \caption{DER Maximization}
  \label{dist_pvc}
\end{subfigure}
\caption{Voltage Distributions}
\label{dist_lp}
\end{figure}

In Fig.~\ref{dist_lp}, we show how the voltages in the system are impacted by the loss minimization (scenario (ii)) and DER maximization (scenario (iii)) objectives respectively. In the loss-minimization case, we expect that all of the bus voltages will be shifted towards the upper limit. In the DER maximization case we expect that bus voltages that exceed the voltage limit will be reduced to be within the voltage limits. These expectations are confirmed in Fig.~\ref{dist_lp}.

\subsubsection{Communication Network Stress Test}
\begin{figure}[t]
\centering
\begin{subfigure}{.24\textwidth}
  \centering
  \includegraphics[width=1.1\linewidth]{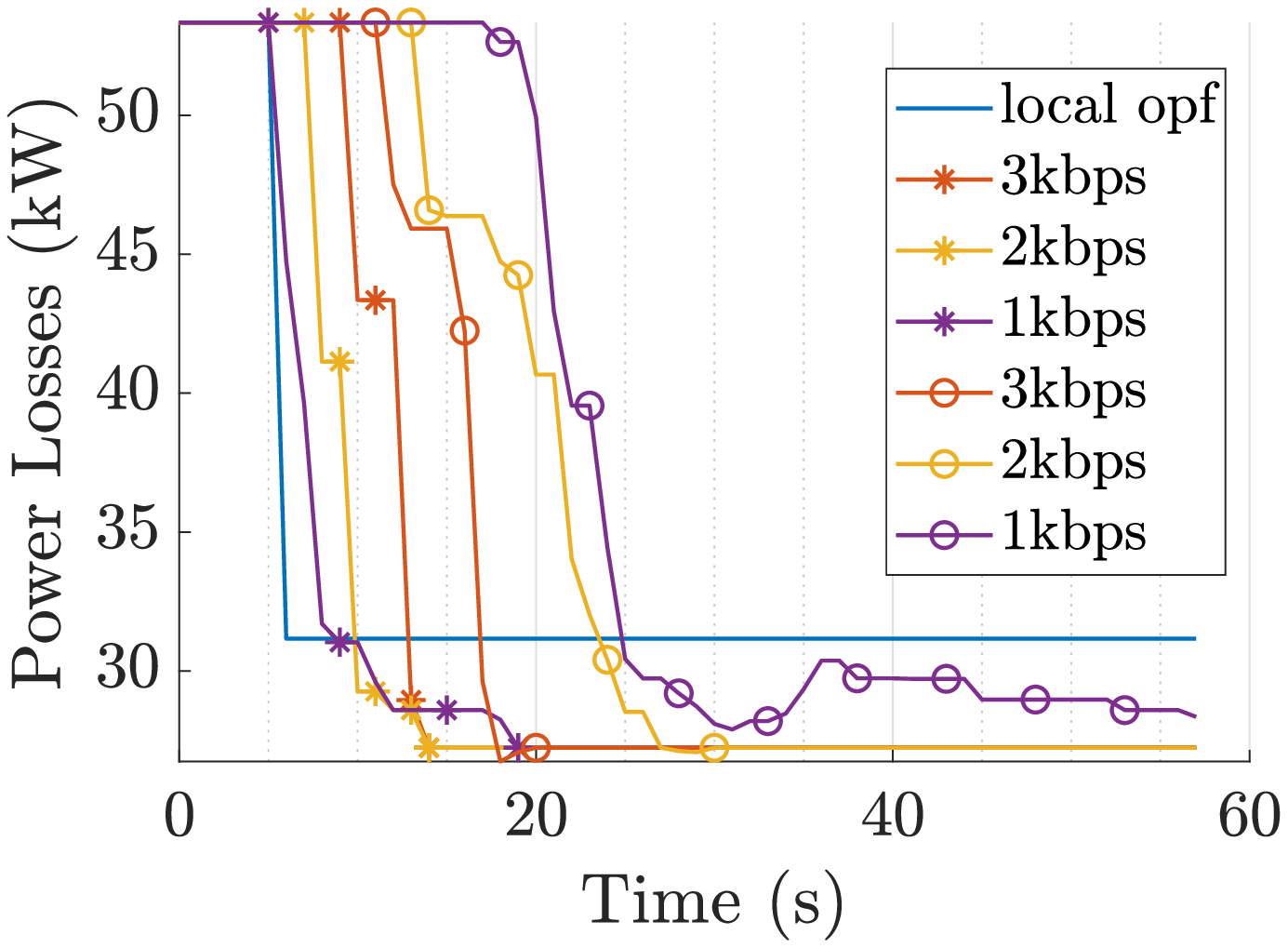}
  \vspace{-0.6cm}
  \caption{Loss Minimization}
  \label{val_lm}
\end{subfigure}
\begin{subfigure}{.24\textwidth}
  \centering
  \includegraphics[width=1.1\linewidth]{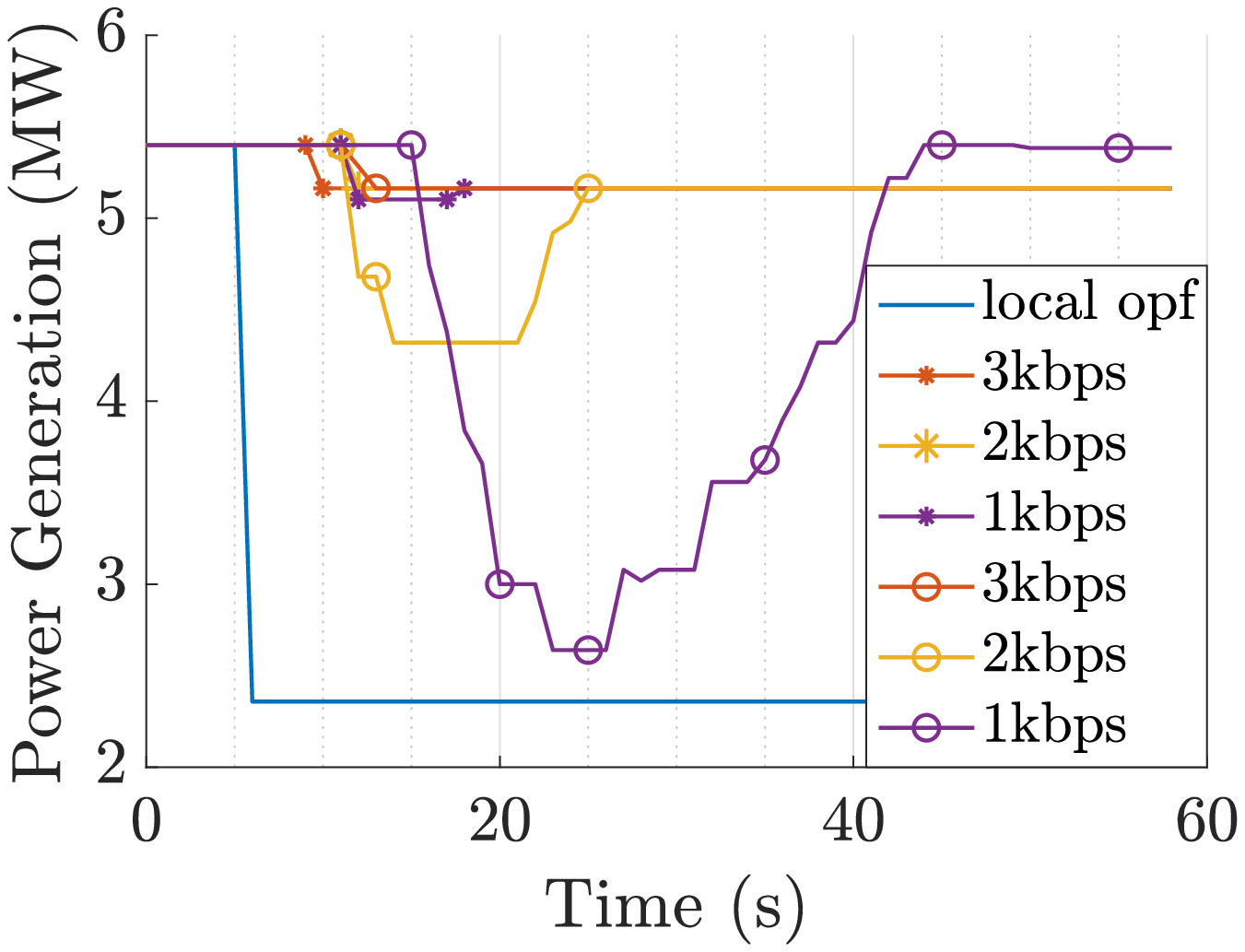}
  \caption{DER Maximization}
  \label{val_pvc}
\end{subfigure}
\caption{Communication Stress Test}
\label{comms_lp}
\end{figure}

Next we test the robustness of the proposed algorithm, when subjected to poor communication network conditions; it is tested with two different communication network topologies (ideal and ring topology) for loss minimization problem \textbf{(D2)} with scenario (ii) and for DER maximization problem \textbf{(D4)} with scenario (iii). In the ideal topology, each area controller has a direct link with its neighbors and with devices in its own area. In the ring topology all devices and controllers are connected in a single large loop without regard to physical location. Each link between network nodes is a point-to-point link with a negligible delay and has different bandwidths (3kbps, 2kbps, or 1kbps) to show the worst-case performance of the algorithm (see Fig. \ref{comms_lp}). The lines marked with circles indicate results run with the ring topology and lines marked with a star indicate the ideal topology. The blue line shows the results if no communication is allowed between controllers and the controllers have perfect communication links with local devices.
There are several effects of poor communications that we have observed:
\begin{itemize}[leftmargin=*, noitemsep]
    \item  If a controller doesn't receive data from either UA or DA because of data delays, it will falsely assume convergence and dispatch the inverters prematurely. When the delayed data does arrive it will continue iterating again.
    \item Delayed communication between areas does not prevent convergence except in extreme cases.
    \item The quality of the resulting solution as measured after convergence is not impacted by the communication delays.
\end{itemize}

\subsubsection{Advantages of Using DT with LP}
\begin{figure}[t]
\centering
\begin{subfigure}{.24\textwidth}
  \centering
  \includegraphics[width=1.1\linewidth]{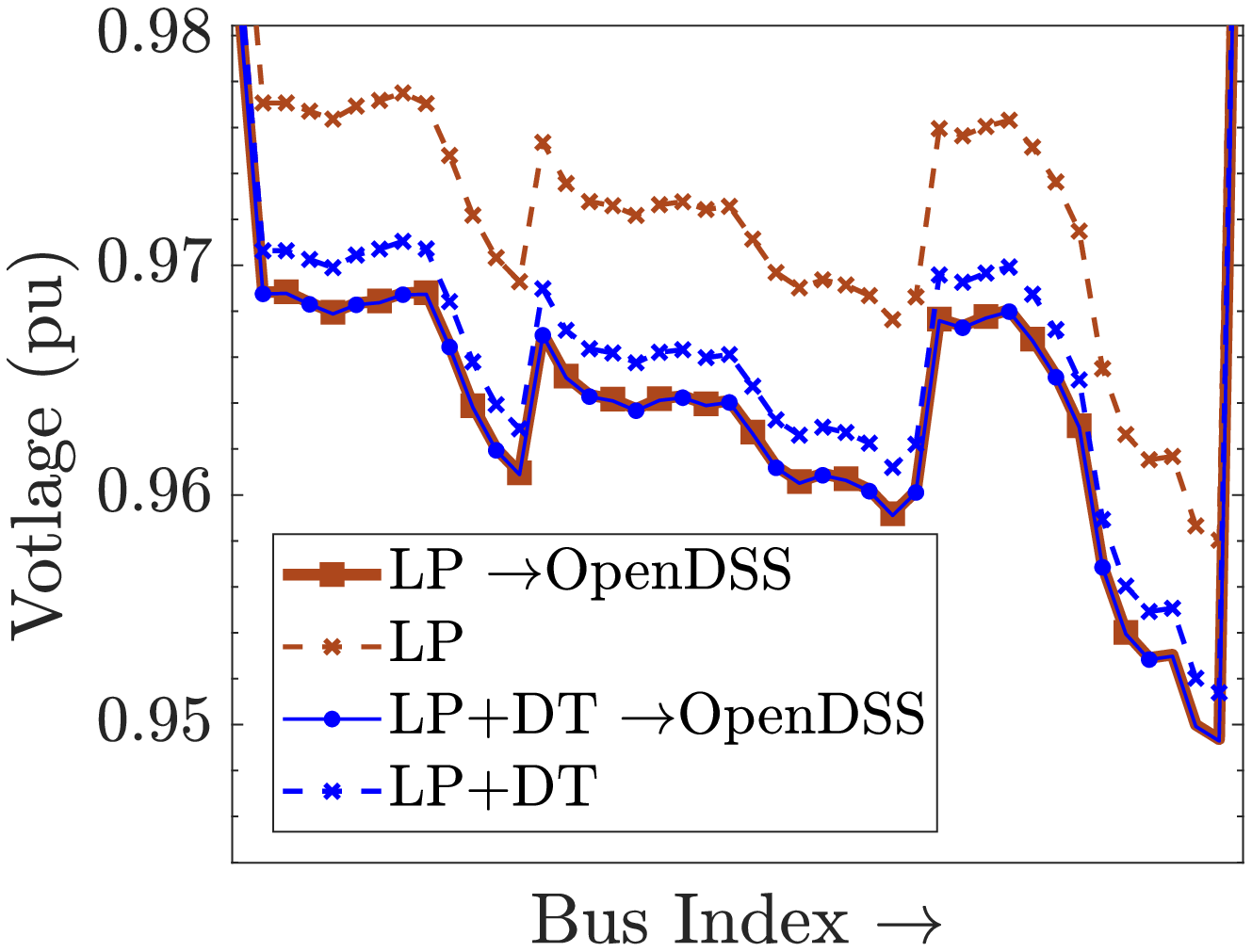}
  \caption{Loss Minimization}
  \label{fig:DT_corner_lm}
\end{subfigure}
\begin{subfigure}{.24\textwidth}
  \centering
  \includegraphics[width=1.1\linewidth]{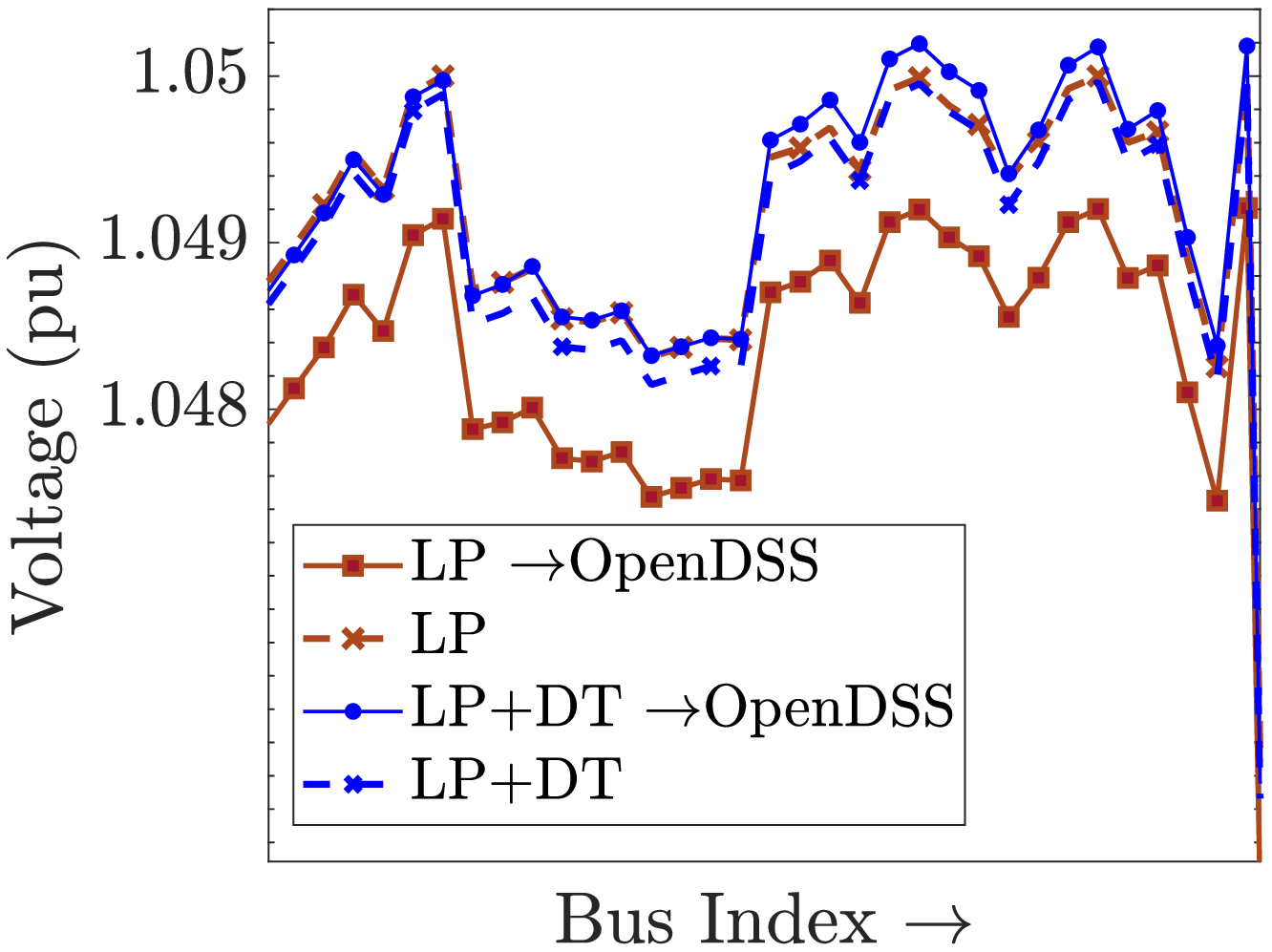}
  \caption{DER Maximization}
  \label{fig:DT_corner_pv}
\end{subfigure}
\caption{Voltage Approximation Error}
\label{fig:DT_corner}
\end{figure}

The LP model can estimate the nodal voltages with moderate accuracy; however, for corner cases, such as when the system is operating at the voltage limits, the approximation error can lead to a higher optimality gap and even infeasible solutions. In Fig. \ref{fig:DT_corner_lm}, we have shown a snippet of voltage at phase C for a high loading condition; using a DT model yields more accurate nonlinear nodal voltages than LP solutions that approximate higher voltages. If the voltage lower bound constraint is active, an overestimation of the voltage can lead to infeasible space upon implementing the LP solutions to the actual system. However, by using the DT model, we can estimate the nonlinear voltages more accurately and can preemptively tackle any infeasibility. Again for the DER maximization case, when the voltage upper bound is active, the higher approximated voltage can lead us to sub-optimal DER generation solutions. Here, a low loading condition scenario has been simulated, and a snippet of phase A voltages are shown in Fig. \ref{fig:DT_corner_pv}. Upon implementing the LP solution in the OpenDSS, we obtain lower feeder voltages indicating that more DER can be integrated before the actual feeder limit is reached. Thus, LP model results in sub-optimal solutions. However, the LP+DT model does not overestimate the voltages and hence optimized for the maximum generation that can be integrated without causing any voltage violations. We observed that the optimal solutions for the two models differ by $300$ kW. This case further demonstrates the usability of the DT model in accurately solving the OPF problems without augmenting any complicated nonlinear formulation, but still resulting in more accurate optimal solutions. 

\vspace{-0.2cm}
\section{Conclusions}
\vspace{-0.05cm}
In this paper, a novel distributed optimal power flow method for radial three-phase, unbalanced power distribution system has been proposed. The proposed approach uses {\textit{simulation within optimization}} to improve the scalability and reduce solution time for each local subproblem. We employ detailed DT models of the local network in conjugation with a linear-approximated OPF to ensure a power-flow feasible and near-optimal solution for multiple OPF problems for power distribution system. Additionally, the proposed distributed coordination method achieves a converged network-level optimal solution using significantly fewer communication rounds. We also evaluated the performance of the proposed algorithm using a cyber-power co-simulation environment with various communication network parameters, to validate the robustness of the proposed distributed OPF algorithm under stressed communication. For various cases, it is shown that the proposed method successfully converges to the network-level optimal solution within reasonable time, even when the communication network is facing significant stress, such as, communication delays and a low bandwidth communication infrastructure.


\vspace{-0.3cm}
\bibliographystyle{ieeetr}
\bibliography{cite}

\begin{thebibliography}{10}

\bibitem{momoh1999review1}
J.~A. Momoh, R.~Adapa, and M.~El-Hawary, ``A review of selected optimal power
  flow literature to 1993. i. nonlinear and quadratic programming approaches,''
  {\em IEEE transactions on power systems}, vol.~14, no.~1, pp.~96--104, 1999.

\bibitem{castillo2013survey}
A.~Castillo and R.~P. O’Neill, ``Survey of approaches to solving the
  {ACOPF},'' {\em US Federal Energy Regulatory Commission, Tech. Rep}, 2013.

\bibitem{molzahn2017survey}
D.~K. Molzahn, F.~D{\"o}rfler, H.~Sandberg, S.~H. Low, S.~Chakrabarti,
  R.~Baldick, and J.~Lavaei, ``A survey of distributed optimization and control
  algorithms for electric power systems,'' {\em IEEE Transactions on Smart
  Grid}, vol.~8, no.~6, pp.~2941--2962, 2017.

\bibitem{kim2000comparison}
B.~H. Kim and R.~Baldick, ``A comparison of distributed optimal power flow
  algorithms,'' {\em IEEE Transactions on Power Systems}, vol.~15, no.~2,
  pp.~599--604, 2000.

\bibitem{zheng2015fully}
W.~Zheng, W.~Wu, B.~Zhang, H.~Sun, and Y.~Liu, ``A fully distributed reactive
  power optimization and control method for active distribution networks,''
  {\em IEEE Transactions on Smart Grid}, vol.~7, no.~2, pp.~1021--1033, 2015.

\bibitem{boyd2011distributed}
S.~Boyd, N.~Parikh, E.~Chu, B.~Peleato, J.~Eckstein, {\em et~al.},
  ``Distributed optimization and statistical learning via the alternating
  direction method of multipliers,'' {\em Foundations and
  Trends{\textregistered} in Machine learning}, vol.~3, no.~1, pp.~1--122,
  2011.

\bibitem{qu2017harnessing}
G.~Qu and N.~Li, ``Harnessing smoothness to accelerate distributed
  optimization,'' {\em IEEE Transactions on Control of Network Systems},
  vol.~5, no.~3, pp.~1245--1260, 2017.

\bibitem{mhanna2018adaptive}
S.~Mhanna, G.~Verbi{\v{c}}, and A.~C. Chapman, ``Adaptive admm for distributed
  ac optimal power flow,'' {\em IEEE Transactions on Power Systems}, vol.~34,
  no.~3, pp.~2025--2035, 2018.

\bibitem{peng2016distributed}
Q.~Peng and S.~H. Low, ``Distributed optimal power flow algorithm for radial
  networks, i: Balanced single phase case,'' {\em IEEE Transactions on Smart
  Grid}, vol.~9, no.~1, pp.~111--121, 2016.

\bibitem{april2004new}
J.~April, M.~Better, F.~Glover, and J.~Kelly, ``New advances and applications
  for marrying simulation and optimization,'' in {\em Proceedings of the 2004
  Winter Simulation Conference, 2004.}, vol.~1, IEEE, 2004.

\bibitem{nakayama2002simulation}
H.~Nakayama, M.~Arakawa, and R.~Sasaki, ``Simulation-based optimization using
  computational intelligence,'' {\em Optimization and Engineering}, vol.~3,
  no.~2, pp.~201--214, 2002.

\bibitem{nguyen2014review}
A.-T. Nguyen, S.~Reiter, and P.~Rigo, ``A review on simulation-based
  optimization methods applied to building performance analysis,'' {\em Applied
  energy}, vol.~113, pp.~1043--1058, 2014.

\bibitem{zhou2019digital}
M.~Zhou, J.~Yan, and D.~Feng, ``Digital twin framework and its application to
  power grid online analysis,'' {\em CSEE Journal of Power and Energy Systems},
  vol.~5, no.~3, pp.~391--398, 2019.

\bibitem{pan2020digital}
H.~Pan, Z.~Dou, Y.~Cai, W.~Li, X.~Lei, and D.~Han, ``Digital twin and its
  application in power system,'' in {\em 2020 5th International Conference on
  Power and Renewable Energy (ICPRE)}, pp.~21--26, IEEE, 2020.

\bibitem{brosinsky2018recent}
C.~Brosinsky, D.~Westermann, and R.~Krebs, ``Recent and prospective
  developments in power system control centers: Adapting the digital twin
  technology for application in power system control centers,'' in {\em 2018
  IEEE International Energy Conference}, pp.~1--6, IEEE, 2018.

\bibitem{he2019preliminary}
X.~He, Q.~Ai, R.~C. Qiu, and D.~Zhang, ``Preliminary exploration on digital
  twin for power systems: Challenges, framework, and applications,'' {\em arXiv
  preprint arXiv:1909.06977}, 2019.

\bibitem{darbali2021state}
R.~Darbali-Zamora, J.~Johnson, A.~Summers, C.~B. Jones, C.~Hansen, and
  C.~Showalter, ``State estimation-based distributed energy resource
  optimization for distribution voltage regulation in telemetry-sparse
  environments using a real-time digital twin,'' {\em Energies}, vol.~14,
  no.~3, p.~774, 2021.

\bibitem{erseghe2014distributed}
T.~Erseghe, ``Distributed optimal power flow using admm,'' {\em IEEE
  transactions on power systems}, vol.~29, no.~5, pp.~2370--2380, 2014.

\bibitem{dall2013distributed}
E.~Dall'Anese, H.~Zhu, and G.~B. Giannakis, ``Distributed optimal power flow
  for smart microgrids,'' {\em IEEE Transactions on Smart Grid}, vol.~4, no.~3,
  pp.~1464--1475, 2013.

\bibitem{bernstein2019real}
A.~Bernstein and E.~Dall’Anese, ``Real-time feedback-based optimization of
  distribution grids: A unified approach,'' {\em IEEE Transactions on Control
  of Network Systems}, vol.~6, no.~3, pp.~1197--1209, 2019.

\bibitem{magnusson2020distributed}
S.~Magn{\'u}sson, G.~Qu, and N.~Li, ``Distributed optimal voltage control with
  asynchronous and delayed communication,'' {\em IEEE Transactions on Smart
  Grid}, 2020.

\bibitem{sadnan2021distributed}
R.~Sadnan and A.~Dubey, ``Distributed optimization using reduced network
  equivalents for radial power distribution systems,'' {\em IEEE Transactions
  on Power Systems}, vol.~36, no.~4, pp.~3645--3656, 2021.

\bibitem{jha2019bi}
R.~R. Jha, A.~Dubey, C.-C. Liu, and K.~P. Schneider, ``Bi-level volt-var
  optimization to coordinate smart inverters with voltage control devices,''
  {\em IEEE Transactions on Power Systems}, vol.~34, no.~3, pp.~1801--1813,
  2019.

\bibitem{sadnan2021online}
R.~Sadnan, T.~Asaki, and A.~Dubey, ``Online distributed optimization in radial
  power distribution systems: Closed-form expressions,'' in {\em 2021 IEEE
  International Conference on Communications, Control, and Computing
  Technologies for Smart Grids (SmartGridComm)}, pp.~51--56, IEEE, 2021.

\bibitem{palmintier2017design}
B.~Palmintier, D.~Krishnamurthy, P.~Top, S.~Smith, J.~Daily, and J.~Fuller,
  ``Design of the helics high-performance
  transmission-distribution-communication-market co-simulation framework,'' in
  {\em 2017 Workshop on Modeling and Simulation of Cyber-Physical Energy
  Systems (MSCPES)}, pp.~1--6, IEEE, 2017.

\bibitem{Chassin:2008fq}
D.~P. Chassin, K.~P. Schneider, and C.~Gerkensmeyer, ``{GridLAB-D: An
  open-source power systems modeling and simulation environment},'' in {\em
  {IEEE/PES Transmission and Distribution Conference and Exposition}},
  pp.~1--5, IEEE, May 2008.

\bibitem{ns3}
{NS-3 Network Simulator}.
\newblock https://www.nsnam.org/.

\end{thebibliography}


\end{document}